\documentclass[a4paper,12pt]{article}

\usepackage{cite}
\usepackage{amsmath}
\usepackage{epsf}
\usepackage{comment}

\title{\textbf{Exclusion of measurements with excessive residuals}}

\begin{document}
\sloppy

\begin{center}

\large{\textbf{Exclusion of measurements with excessive residuals 
(blunders) in estimating model parameters}}\\
\bigskip
\thispagestyle{empty}

\normalsize
{I.~I.~NIKIFOROV*}\\
\bigskip

\small

Sobolev Astronomical Institute of St.~Petersburg State University,\\ 
Universitetskij Prospekt 28, 
Staryj Peterhof, St.~Petersburg 198504, Russia\\
\smallskip

*Email: nii@astro.spbu.ru\\

\end{center}

\footnotesize

An adjustable algorithm of exclusion of conditional equations with  excessive
residuals is proposed. The criteria applied in the algorithm use variable
exclusion limits which decrease as the number of equations goes down. The
algorithm is easy to use, it possesses rapid convergence, minimal
subjectivity, and high degree of generality.
\smallskip


\normalsize

\section*{\normalsize\textbf{1. Introduction}}

In many astronomical (and not only astronomical) problems of estimation of model
parameters, it is important reasonably to exclude unreliable data which produce
large residuals, i.e., deviations, $\varepsilon$, of measurements from the
accepted model: $|\varepsilon_j|/\sigma_j\gg 1$, where $\sigma_j$ is the
standard deviation for $j$-th measurement, $j=1,\:\ldots,\:N$, $N$ is the
number of measurements, i.e., of conditional equations. The occurrence of large
residuals (``blunders'') contradicts the basic assumption of least-squares fitting
on the normal distribution of measurement errors and can cause strong biases
of parameter estimates. The common ``$3\sigma$'' criterion to exclude blunders
\begin{equation}
\label{3s}      
\frac{|\varepsilon_j|}{\sigma_j}>k=3
\end{equation}
does not allow for the probability of accidental occurrence of residual
(\ref{3s}) to increase with $N$ and become not negligible already at $N$ of
order several tens.

In this paper, a more adjustable algorithm of exclusion of equations with
excessive residuals on the basis of a variable criterion limit is elaborated.

\section*{\normalsize\textbf{2.\kern0.5emAlgorithm of excluding
measurements with excessive residuals}}

1.\kern0.5emFor a given $N$, a value of $\kappa$ which satisfies the equation 
\begin{equation}       
\left[1-\psi( \kappa)\right]N=1, \qquad  \psi(z)\equiv\sqrt{\frac2\pi}\int^z_0 e^{-\frac{1}{2}t^2}dt,
\end{equation}       
where $\psi(z)$ is the probability integral, is found.
The expectation value for the number of conditional equations with residuals
\begin{equation}
\label{k1s}       
{|\varepsilon_j|/\sigma_j}>\kappa,
\end{equation}
equals one, if residuals are normally distributed. A larger number of
equations with such residuals may be considered as probably excessive.

2.\kern0.5emThe number $L$ of equations satisfying the criterion (\ref{k1s})
is determined. 

3.\kern0.5emIf $L>1$, $L-L'$ equations with the largest values of
$|\varepsilon_j|/\sigma_j$ are excluded from consideration. Here, $L'\ge 1$ is
a parameter of the algorithm.  


4.\kern0.5emThe criterion (\ref{3s}) with $k$ depending on $N$ is applied to
the remaining equations, in particular if $L=1$:
\begin{equation}\label{kg}
      {|\varepsilon_j|/\sigma_j}>k_\gamma(N),
\end{equation}
where $k_\gamma$ is the root of the equation
\begin{equation}\label{kg(N)}
 1-\left[\psi( k_\gamma)\right]^N=\gamma.
\end{equation}
Here, $\gamma$ is an accepted confidence level. 
For low $\gamma$, i.e., for low $1-\psi(k_\gamma)$, in lieu of (\ref{kg(N)}) 
an approximate equation can be used:
\begin{equation}\label{kg(N)1}
      [1-\psi( k_\gamma)]N=\gamma.
\end{equation}    

 
5.\kern0.5emFollowing the exclusion of equations with excessive residuals, a
new solution of the problem is found from the remaining equations. Thereupon
points 1--4 of this algorithm are applied again with new estimates of
parameters and $\sigma_j$. The iterations are interrupted if no further
exclusion happens.

The probability ${\cal P}(L)$ of accidental occurrence of $L$ residuals
satisfying (\ref{k1s}) can be approximately evaluated with the Poisson
distribution, which is ${\cal P}(L)={e^{-1}/L!}$\, in this case. 
This approach gives
     $$
      {\cal P}(L\ge 2)  \approx  0.264,\qquad
      {\cal P}(L\ge 3)  \approx   0.080,\qquad
      {\cal P}(L\ge 4)  \approx   0.019.
     $$
Thus numbers of $L=3$ and $4$ can be considered as excessive, i.e., 
$L'=2$ or  3 can be correspondingly accepted. However, if 
unbiased parameters are more important 
than an unbiased residual variance, $L'=1$ is also allowed.

Point 4 of the algorithm is essential in the case of only a single (or few)
very large blunder(s), when point~3 can not come into action. A level of
$\gamma=0.05$, being the standard one in many statistical criteria, can
be accepted.

\section*{\normalsize\textbf{Acknowledgments}}
The work is partly supported by the Russian Foundation for Basic Research
grant 08-02-00361 and the Russian Pre\-si\-dent Grant for State Support of Leading
Scientific Schools of Russia no.\ NSh-1323.2008.2.

\end{document}